\documentclass[12pt]{article}
\usepackage{amssymb}
\begin{document}
\title{{\bf A Nonliearly Dispersive Fifth Order Integrable Equation
    and its Hierarchy}}
\author{ Ashok Das$^{a}$ and Ziemowit Popowicz$^{b}$\\
\\
$^{a}$ Department of Physics and Astronomy\\
University of Rochester\\
Rochester, NY 14627 - 0171, USA\\
\\
$^{b}$ Institute of Theoretical Physics\\
University of Wroc\l aw\\
pl. M. Borna 9, 50 -205 Wroc\l aw, Poland\\ }

\maketitle

\begin{center}
{\bf Abstract}
\end{center}

In this paper, we study the properties of a nonlinearly
dispersive integrable system of fifth order and its associated
hierarchy. We describe a Lax representation for such a system 
which leads to two infinite series of conserved charges and two
hierarchies of equations that share the same conserved charges. We
construct two compatible Hamiltonian structures as well as
their Casimir functionals. One of the structures has a single Casimir
functional while the other has two. This  allows us to extend the
flows into negative order and clarifies the meaning of two different
hierarchies of positive flows. We study the behavior of these
systems under a hodograph transformation and show that they are
related to the Kaup-Kupershmidt and the Sawada-Kotera
equations under appropriate Miura transformations. We also
discuss briefly some properties associated with the generalization
of second, third and fourth order Lax operators.

\newpage

\section{Introduction}

There has been a lot of interest in recent years in the study of the
Harry Dym \cite{kruskal,blaszak,brunelli,brunelli1} and the
Hunter-Zheng equations \cite{hunter,camasa} which define an integrable
hierarchy. Besides having applications in various physical systems,
these equations represent a class of equations where the hierarchy can
be extended to both positive as well as negative integer
flows. Furthermore, in these systems of equations, the dispersive
term in the equation is nonlinear, unlike the KdV
equation where the dispersive term is linear in the dynamical
variable. This is manifest in the structure of the Lax operator and
the equation for the Harry Dym equation
\begin{equation}
L = u^{2}\partial^{2},\quad \frac{\partial L}{\partial t} =
\left[\left(L^{3}\right)_{\geq 2}, L\right],\quad u_{t} = u^{3}
u_{xxx}.\label{harrydym}
\end{equation}
In trying to understand the properties of other systems that belong to
this class of nonlinearly dispersive equations, we have chosen to 
consider  the generalized Lax operator for the Harry Dym equation, 
introduced in  \cite{oevel} and study the
resulting systems. This leads us to a very interesting integrable
hierarchy which was proposed a few years ago \cite{holm}, but whose
properties 
were never studied in detail. However, having a Lax representation for
such a system, we are able to analyze various aspects of this system
of equations. In particular, we construct the conserved charges for
the positive flows directly from the Lax operator itself and show that
the system, in fact, contains two infinite families of conserved
charges. Correspondingly, the Lax equation really gives rise to two
hierarchies of equations. Such a behavior has been noted earlier in
connection with dispersionless polytropic gas dynamics \cite{alex},
but to the best of our knowledge, this is the first example of such a
behavior in systems with dispersive terms. We obtain
the two compatible Hamiltonian structures thereby showing that the two
hierarchies are bi-Hamiltonian. We determine the Casimir functionals of both
the Hamiltonian structures which allows us to extend the flows to
negative orders (which are nonlocal). We construct the recursion
operator as well as its inverse which allows us to construct the
non-local charges and the nonlocal flows of the system recursively. We
also study the behavior of these systems under a hodograph
transformation and make connection with the Kaup-Kupershmidt and
Sawada-Kotera equations.

The paper is organized as follows. In section {\bf 2}, we discuss the
Lax representation and the two associated hierarchies of integrable
equations. In section {\bf 3} we present the conserved charges
following from the Lax operator as well as the  two compatible
Hamiltonian structures which also guarantees integrability of the
system. The Casimir functionals of the system are obtained which
allows us to extend the flows to negative orders. We construct
explicitly the first few conserved nonlocal charges through the
recursion operator. In section {\bf 4}, we study the behavior of these
systems under a hodograph transformation and make connection with the
Kaup-Kupershmidt and the Sawada-Kotera equations. We point out
various interesting properties of the systems under the hodograph
transformation. We discuss briefly, in sec {\bf 5}, some features
associated with a 
generalized Lax operator of second, third and the fourth order and
conclude with a brief summary in section {\bf 6}.

\section{Lax Representation}

Let us consider  the Lax operator introduced by Konopelchenko and Oevel \cite{oevel}
\begin{equation}
L=u^3\partial^{3}.\label{generalization}
\end{equation}
It can be checked with some algebra that this Lax operator leads to
two consistent  hierarchies of equations of the form
\begin{equation}
\frac{\partial L}{\partial t^{(\mp)}_n} = 
\left[\left(L^{2n \mp \frac{1}{3}}\right)_{\geq
    2},L\right],\label{newlaxequation}
\end{equation}
where $ n=1,2,3\ldots$. Let us note that the first few equations of
the two hierarchies have the forms
\begin{eqnarray}
u_{t^{(-)}_1} &=& u^4 \big(u^2 \big)_{5x} \nonumber\\
\noalign{\vskip 4pt}%
u_{t_{2}^{(-)}} & = & -\frac{1}{81}\,u^{4}\left[30 u^{7} u_{6x} + 270
  u^{6} u_{x} u_{5x}\cdots\right]_{5x},\nonumber\\
\noalign{\vskip 4pt}%
u_{t^{(+)}_1} &=& \frac{1}{3} u^4 \big(2u^{3} u_{xx} - u^{2}
u_{x}^{2}\big)_{5x},\nonumber\\
\noalign{\vskip 4pt}%
u_{t_{2}^{(+)}} & = & -\frac{1}{243}\,u^{4}\left[42 u^{9} u_{8x} + 840
  u^{8}u_{x}u_{7x}\cdots \right]_{5x}.\label{newequations} 
\end{eqnarray}
These are even more nonlinearly dispersive than
the Harry Dym equation (as we had anticipated earlier) and we note that
in terms of the variable $\varphi = u^{-3}$, the two lowest order
equations take the forms
\begin{eqnarray}
\varphi_{t^{(-)}_1} & = & \big(\varphi^{-2/3} \big)_{5x}\nonumber\\
\noalign{\vskip 4pt}%
\varphi_{t^{(+)}_1} & = & -2 \big(\varphi^{-1} \big(
\big(\varphi^{-1/3} \big)_{xx} - 
2\big(\varphi^{-1/6} \big)^2_x\big)\Big)_{5x}. 
\end{eqnarray}
These equations indeed coincide with the equations considered by Holm
and Qiao \cite{holm}. However, having a Lax representation for these
two 
hierarchies, we can study their properties systematically in the
following. In particular, the Lax representation shows that the two
equations belong to two distinct hierarchies (this will become more
clear in the next section) and guarantees that the two hierarchies are
integrable. 

We recall here that the Lax operator for the Harry Dym equation is 
unique. In fact, it can be checked that the only generalization of the
Lax operator (\ref{harrydym}) that leads to a consistent equation has
the form
\begin{eqnarray}
L &=& u^{2} \partial^{2} + uu_{x} \partial = (u\partial)^2, \nonumber \\
\frac{\partial L}{\partial t} &=& \left[\left(L^{3/2}\right)_{\geq 2}, L\right],
\end{eqnarray}
and leads to the equation
\begin{equation}
u_{t} = \left(u^{3}u_{xx}\right)_{x},\label{thirdorder}
\end{equation}
which is different from (\ref{harrydym}). However, in the present
case, we  find that the Lax operator is not
unique. For example, it can
be checked that the Lax operator and the equation
\begin{equation}
L = u^{3} \partial^{3} + \frac{3}{2} u^{2}u_{x} \partial^{2},\quad
\frac{\partial L}{\partial t^{(-)}_{1}} =
\left[\left(L^{5/3}\right)_{\geq 2}, L\right],\label{secondlax}
\end{equation}
also lead to the fifth order equation in (\ref{newequations}). We will
comment on the general third order Lax operator in one dynamical
variable in sec {\bf 5}.

Finally, we note that one can try to generalize such Lax operators to
even higher orders. However, a simple Lax operator and equation such as
\begin{equation}
L = u^{m} \partial^{m},\quad m\geq 4,\quad \frac{\partial L}{\partial
  t} = \left[\left(L^{n/m}\right)_{\geq 2}, L\right],
\end{equation}
where $n$ is not a multiple of $m$, does not lead to consistent equations.

\section{Bi-Hamiltonian Structure}

In this section we will show that the equations (\ref{newequations})
are Hamiltonian equations. In fact, as we will show, they represent a
bi-Hamiltonian system of equations. From the structure of the Lax
operator, we note that we can define two series of conserved charges
for the system corresponding to (up to multiplicative constants) 
\begin{equation}
H_{n}^{(-)} = \int \mathrm{d}x\,{\rm Tr}\,L^{2n-\frac{1}{3}},\quad
H_{n}^{(+)} = \int \mathrm{d}x\,{\rm Tr}\,L^{2n +
  \frac{1}{3}},\quad n=0,1,2,\ldots,\label{charges}
\end{equation}
where ``Tr'' represents the Adler trace over pseudo-differential
operators. 
The other fractional powers of the Lax operator can be easily shown to
give trivial charges. Even though the charges in (\ref{charges}) come
from the same Lax operator, we have written them as two distinct
series for reasons that will become clear shortly. We note here that
the first few charges of the series can be written explicitly as
\begin{eqnarray}
H_{0}^{(-)} & = & -\int \mathrm{d}x\,\frac{1}{u},\nonumber\\
\noalign{\vskip 4pt}%
H_{1}^{(-)} & = & \frac{1}{81}\int \mathrm{d}x\, \big(-6u_{6x}u^4
-54u_{5x}u_xu^3 -102u_{4x}u_{xx}u^3 - 
84u_{4x}u_{x}^2u^2 -57u_{3x}^2u^3 \nonumber\\  
&& \quad - 204u_{3x}u_{2x}u_xu^2 + 12u_{3x}u_x^3u -40u_{2x}^3u^2 + 60u_{xx}^2
u - 30u_{xx}u_x^4 + 5 u_x^6u^{-1} \big),\nonumber\\
\noalign{\vskip 4pt}%
H_{0}^{(+)} & = & - \frac{1}{3} \int
\mathrm{d}x\,\frac{u_{x}^{2}}{u},\nonumber\\ 
\noalign{\vskip 4pt}%
H_{1}^{(+)} & = &\!\!- \frac{1}{243}\int \mathrm{d}x\left[6u^{6}u_{8x} + 120
  u^{5}u_{x}u_{7x} + 264 u^{5}u_{xx}u_{6x} + 372 u^{5}u_{xxx}u_{5x} +
  213 u^{5} u_{4x}^{2}\right.\nonumber\\
\noalign{\vskip 4pt}%
 &  &  \quad + 708 u^{4} u_{x}^{2} u_{6x} + 2376 u^{4} u_{x}u_{xx}
u_{5x} + 2712 u^{4} u_{x} u_{xxx}u_{4x} + 1488 u^{4} u_{xx}^{2}
u_{4x}\nonumber\\ 
\noalign{\vskip 4pt}%
 &  & \quad + 1308 u^{4}u_{xx}u_{xxx}^{2} + 1332 u^{3} u_{x}^{3}
u_{5x} + 4452 u^{3} u_{x}^{2} u_{xx}u_{4x} + 2058 u^{3} u_{x}^{2}
u_{xxx}^{2} \nonumber\\
\noalign{\vskip 4pt}%
 &  & \quad + 2976 u^{3}u_{x}u_{xx}^{2} u_{xxx} + 112 u^{3}u_{xx}^{4}
+ 552 u^{2}u_{x}^{4} u_{4x} + 984 u^{2}u_{x}^{2} u_{xxx}^{2} \nonumber\\
\noalign{\vskip 4pt}%
 &  & \quad\left.  - 224
u^{2} u_{x}^{2} u_{xx}^{3} + 24 u
u_{x}^{5} u_{xxx} + 168 u u_{x}^{4} u_{xx}^{2} - 56 u_{x}^{6} u_{xx} +
7u^{-1} u_{x}^{8}\right]. \label{charges1} 
\end{eqnarray}
In addition, we have found another charge (not following from the Lax
operator) that is also conserved,
\begin{equation}
H_{-1} = \int \mathrm{d}x\,\frac{1}{u^{3}}.\label{charges2}
\end{equation}

Given the conserved charges in (\ref{charges1}) and (\ref{charges2}),
it is now easy to check that the two hierarchies of nonlinearly 
dispersive equations in
(\ref{newequations}) can be written in the bi-Hamiltonian form
\begin{eqnarray}
u_{t_{n}^{(-)}} & = & {\cal D}_1 \frac{\delta H_{n-1}^{(-)}}{\delta u} =
{\cal D}_2 \frac{\delta H_{n}^{(-)}}{\delta u}, \nonumber\\
\noalign{\vskip 4pt}%
u_{t_{n}^{(+)}} & = & {\cal D}_{1} \frac{\delta H_{n-1}^{(+)}}{\delta u}
  = {\cal D}_{2} \frac{\delta H_{n}^{(+)}}{\delta u},\quad n=1,2,\ldots,
\end{eqnarray}
where 
\begin{eqnarray}
{\cal D}_1 &=&  u^4\partial^5u^4 \nonumber\\
\noalign{\vskip 4pt}%
{\cal D}_2 &=& u^2 \partial u^{-1}\partial^{-3} u^{-1} \partial
u^2.\label{hamiltonianstructure}
\end{eqnarray}
The Hamiltonian structures in (\ref{hamiltonianstructure}) are
manifestly anti-symmetric. It can also be checked using the method of
prolongation \cite{olver} that they satisfy Jacobi identity. For
example, for the 
first structure, we have
\begin{eqnarray}
{\cal D}_{1} \theta & = & u^{4}
\left(u^{4}\theta\right)_{5x},\nonumber\\
\noalign{\vskip 4pt}%
\Theta_{{\cal D}_{1}} & = & \frac{1}{2} \int \mathrm{d}x\,
\left(u^{4}\theta\right)\wedge \left(u^{4}\theta\right)_{5x},\nonumber\\
\noalign{\vskip 4pt}%
{\rm\bf pr}\,V_{{\cal D}_{1}\theta} \left(\Theta_{{\cal D}_{1}}\right)
& = & 4 \int \mathrm{d}x\,\left(u^{4}\theta\right)_{5x}\wedge
\left(u^{7}\theta\right)\wedge \left(u^{4}\theta\right)_{5x} = 0.
\end{eqnarray}
Similarly, the vanishing of the prolongation for the second structure
can also be
easily checked. The compatibility of the two Hamiltonian structures in
(\ref{hamiltonianstructure}) can also be checked through the method of
prolongation \cite{olver} with a little bit of algebra. This shows that
the two hierarchies  
of equations (\ref{newequations}) are indeed bi-Hamiltonian. As a
result, they are integrable.

We note that because of the simplicity in the structure in
(\ref{hamiltonianstructure}), we can construct the recursion operator
for the system in the closed form as
\begin{equation}
{\cal R} = {\cal D}_{2}^{-1} {\cal D}_{1} = u^{-2}\partial^{-1} u
\partial^{3} u \partial^{-1} u^{2} \partial^{5}
u^{4},\label{recursion}
\end{equation}
and it can be checked that the conserved charges (\ref{charges1}) are
related recursively as
\begin{equation}
\frac{\delta H_{n+1}^{(\mp)}}{\delta u} = {\cal R}\,\frac{\delta
  H_{n}^{(\mp)}}{\delta u},\quad
  n=0,1,2,\ldots.\label{recursionrelation} 
\end{equation}
We note also that both of the Hamiltonian structures in
(\ref{hamiltonianstructure}) have Casimir functionals corresponding to 
conserved quantities whose gradients are annihilated by the appropriate
Hamiltonian structure. Thus, for example, ${\cal D}_{1}$ has the
Casimir functional $H_{-1}$ defined in (\ref{charges2}) such that
\begin{equation}
{\cal D}_{1}\,\frac{\delta H_{-1}}{\delta u} = 0.\label{casimir1}
\end{equation}
Similarly, ${\cal D}_{2}$ has two Casimir functionals, namely,
$H_{0}^{(-)}, H_{0}^{(+)}$ such that
\begin{equation}
{\cal D}_{2}\,\frac{\delta H_{0}^{(\mp)}}{\delta u} = 0
.\label{casimir2} 
\end{equation}
This as well as (\ref{recursionrelation}) clarifies why we grouped the
conserved charges into two infinite series even though they arise from
the same Lax operator.

Because of the existence of Casimir functionals, it is possible to
extend the flows into negative orders. In fact, let us note that the
recursion operator (\ref{recursion}) can be inverted in a closed form
to give
\begin{equation}
{\cal R}^{-1} = {\cal D}_{1}^{-1} {\cal D}_{2} = u^{-4} \partial^{-5}
u^{-2} \partial u^{-1}
\partial^{-3} u^{-1} \partial u^{2},\label{inverserecursion}
\end{equation}
and we can define negative flows for negative values of $n$ through
the recursion relation
\begin{equation}
\frac{\delta H_{-n-1}}{\delta u} = {\cal R}^{-1}\,\frac{\delta
H_{-n}}{\delta u},\quad n=1,2,\ldots.
\end{equation}
Using these one can construct the charges associated with the negative
flows recursively and the first few take the forms
\begin{eqnarray}
H_{-1} & = & \int \mathrm{d}x\,\frac{1}{u^{3}},\nonumber\\
\noalign{\vskip 4pt}%
H_{-2} & = & \frac{1}{18} \int
\mathrm{d}x\,\left(\partial^{-2}u^{-3}\right)^{3},\nonumber\\
\noalign{\vskip 4pt}%
H_{-3} & = & \!\!\int \mathrm{d}x\left(\!9
\left(\partial^{-2}u^{-3}\right)
\left(\partial^{-1}\left(\partial^{-2}u^{-3}\right)^{2}\right)^{2}\!\! + 4
\left(\partial^{-2}u^{-3}\right)^{3}
\left(\partial^{-2}\left(\partial^{-2}u^{-3}\right)^{2}\right)\!\!\right),
\label{nonlocalcharges}
\end{eqnarray}
and so on. These nonlocal charges are conserved under the positive
flows by construction, which can also be explicitly checked. However,
at this time we do not know how to obtain them from the Lax operator.

The nonlocal Hamiltonians (\ref{nonlocalcharges}) lead to nonlocal
flows of the form
\begin{equation}
u_{t_{-n}} = {\cal D}_{1}\,\frac{\delta H_{-n-1}}{\delta u} = {\cal
D}_{2}\,\frac{\delta H_{-n}}{\delta u},\quad n=1,2,\cdots.
\end{equation}
The first few flows of the negative order have the explicit forms
\begin{eqnarray}
u_{t_{-1}} & = & - 9 u^{4}
\left(\left(\partial^{-2}u^{-3}\right)^{2}\right)_{xxx},\nonumber\\
\noalign{\vskip 4pt}%
u_{t_{-2}} & = & u^{4}\left[\frac{3}{2}
\left(\partial^{-2}u^{-3}\right)^{2}\partial^{-1}
\left(\partial^{-2}u^{-3}\right)^{2} +
\left(\left(\partial^{-2}u^{-3}\right)^{2} \partial^{-2}
\left(\partial^{-2}u^{-3}\right)^{2}\right)_{x}\right.\nonumber\\
\noalign{\vskip 4pt}%
 & + & \left. \frac{2}{3}
\left(\left(\partial^{-2}u^{-3}\right)\partial^{-2}\left(\partial^{-2}u^{-3}
\right)^{3}\right)_{x} - 3\left(\left(\partial^{-2}u^{-3}\right)
\partial^{-1} \left(\partial^{-2}u^{-3}\right)\partial^{-1}
\left(\partial^{-2}u^{-3}\right)^{2}\right)_{x}\right]_{xx},
\label{nonlocalflows}
\end{eqnarray}
and so on. We note here that if we introduce the variable $ v_{xx} =
u^{-3}$,  then the first equation in (\ref{nonlocalflows}) can also
be rewritten as
\begin{equation}
v_{t_{-2}} = 54vv_x,  
\end{equation}
which is the Riemann equation.

The reason for the existence of two hierarchies of positive flows
(\ref{newequations})  is now clear. As observed earlier
\cite{brunelli2}, because ${\cal D}_{2}$ has two Casimir functionals
(as opposed to one for ${\cal D}_{1}$), a single hierarchy of negative
flows splits up into two branches of positive flows. This in itself is
an extremely interesting point. For, it may suggest a mechanism for a
Lax  description for several negative flows associated with other models.

\section{Hodograph Transformation}

In this section, we will study the behavior of the fifth
order dispersive equation in (\ref{newequations}) under a hodograph
transformation \cite{hodograph}. Let us consider  the change of
variables   
\begin{equation}
x=p(y,\tau),\quad u = p_{y},\quad  \tau = t.\label{transformation}
\end{equation}
Under such a transformation we have 
\begin{eqnarray}
\frac{\partial }{\partial x} & = & \frac{\partial y }{\partial x}
\frac{\partial }{\partial y} = \frac{1}{p_y}\frac{\partial }{\partial
  y} = \frac{1}{u}\,\partial_{y},\nonumber\\
\noalign{\vskip 4pt}%
 \frac{\partial }{\partial \tau} & = & \frac{\partial }{\partial t} +
\frac{p_{\tau}}{p_{y}}\frac{\partial }{\partial y}.
\end{eqnarray} 
Using this as well as the identification $ u=p_y$, it is
straightforward to see that the fifth order equation in
(\ref{newequations}) goes into 
\begin{equation}
p_{\tau} = p_{5y} - 5\, \frac{p_{4y}p_{yy}}{p_y} +
5\,\frac{p_{3y}p_{yy}^2}{p_y^2}. \label{pequation}
\end{equation}
Introducing the variable $f=\frac{p_{yy}}{p_y}$, we can write
(\ref{pequation}) as
\begin{equation}
f_{\tau} = \left(f_{4y} + 5 f_{y} f_{yy} - 5 f^{2} f_{yy} - f
 f_{y}^{2} + f^{5}\right)_{y}.
\end{equation}
This is exactly the equation considered by Fordy and Gibbons
 \cite{fordy}  who also
 showed that by a proper choice of the Miura transformation, it is
 possible to transform this equation to Kaup-Kupershmidt
 \cite{kaup,fordy1}  or
 Sawada-Kotera \cite{sawada} equations. Indeed, if we choose the Miura
 transformation   
\begin{equation}
g = f_y -\frac{1}{2} f^2,
\end{equation}
then (\ref{pequation}) takes the form
\begin{equation}
g_{\tau} = \frac{1}{6} \left(6 g_{4y} + 60 gg_{yy} + 45 g_{y}^{2} + 40
g^{3}\right)_{y},\label{kkequation}
\end{equation}
which is the Kaup-Kupershmidt equation. On the other hand, a slightly
modified Miura transformation of the form
\begin{equation}
g = -f_{y} + f^{2},
\end{equation}
takes (\ref{pequation}) to
\begin{equation}
g_{\tau}=( g_{4y} + 5gg_{yy} + \frac{5}{3}g^3 )_y
\end{equation}
This is, in fact, the Sawada-Kotera equation. This shows that the
fifth order dispersive equation in (\ref{newequations}) can be mapped
into both Kaup-Kupershmidt and Sawada-Kotera equations under a
hodograph transformation depending on the Miura transformation used.

This can be partly seen at the level of the Lax operator also. For example,
we note that under the transformations (\ref{transformation}), the Lax
operator in (\ref{generalization}) transforms to
\begin{equation}
\hat L = \partial_{y}^{3} - 3f\partial_{y}^{2} -(f_y-2f^2)\partial_y,
\end{equation}
where we have used the identification $ f=\frac{p_{yy}}{p_y}$. 
The transformed Lax operator generates the following fifth order
equation
\begin{eqnarray}
\frac{\partial \hat L}{\partial t} &=& 9  \left[\left(\hat
  L^{5/3}\right)_{\geq 1} ,
  L\right]  \nonumber\\
\noalign{\vskip 4pt}%
\Longrightarrow\;f_t & = & (-f_{4y} - 5f_{yy}f_y + 5f_{yy}f^2 +
  5f_y^2f - f^5)_y.
\end{eqnarray}
A gauge transformation takes us to a new Lax operator of the form
\begin{equation}
\bar L = e^{-\int f} \hat L e^{\int f} = \partial_{y}^{3} +
2g\partial_y + g_y, 
\end{equation}
where $ g=f_y - \frac{1}{2}f^2 $ defines the Miura transformation and
we note that this is indeed the Lax operator considered in
\cite{fordy1}. In
terms of this transformed Lax operator, the equation 
\begin{equation}
\frac{\partial \bar{L}}{\partial t} =
9\left[\left(L^{5/3}\right)_{+},\bar{L}\right],
\end{equation}
leads to
\begin{equation}
g_t= \frac{1}{6}\left(6g_{4y} + 60gg_{yy} + 45 g_y^2 + 40 g^3\right)_y.
\end{equation}
We recognize this to be the Kaup-Kupershmidt equation
(\ref{kkequation}). However, starting from (\ref{generalization}), we
have been unable to obtain the
Lax operator for the Sawada-Kotera equation using the hodograph
transformation. We would like to point out that a Lax representation
for the Sawada-Kotera equation is already known \cite{fordy1} and has
the form
\begin{equation}
L = \partial^{3} + g \partial.
\end{equation}

Interestingly enough, under a hodograph transformation the seventh
order equation in (\ref{newequations}) goes over to the next higher
equation of the Kaup-Kupershmidt or the Sawada-Kotera equation
(depending on the Miura transformation used). This is interesting in
that two distinct hierarchies associated with the system of equations
(\ref{newequations}) map onto a single hierarchy of equations under a
hodograph transformation. Finally, to end this section, we note that
one can carry out the hodograph transformation for the negative order
flows in (\ref{nonlocalflows}) as well and we simply point out that
the first of these transforms under a hodograph transformation to the
Liouville equation while ythe second has the form 
\begin{equation}
p_{\tau} = 6v\partial^{-1} p_y\partial^{-1} p_yv^2 + 2\partial^{-1} p_y\partial^{-1}p_yv^3 -
9\partial^{-1}p_yv\partial^{-1}p_yv^2
\end{equation}
where $ v = \partial^{-1}p_y\partial^{-1} p_y^{-2}$.

Finally let us mention that Kawamoto \cite{kawat} has considered quite 
different fifth order equation then our equation 
\begin{equation}
r_t=r^5r_{xxxxx} +5r^4r_xr_{xxxx} + \frac{5}{2}r^4r_{xx}r_{xxx}+ 
\frac{15}{4}r^3r_x^2r_{xxx}
\end{equation}
and used different trsansformation of dependend and nodependent variables 
\begin{equation}
\frac{\partial }{\partial x} \rightarrow \frac{1}{r}
\frac{\partial}{\partial y} \quad , \quad  \frac{\partial }{\partial t} \rightarrow 
\frac{\partial }{\partial \tau} - r_x\frac{\partial }{\partial y}
 \end{equation}
 in order to put forward the connection of the equation (41) with the Kaup - Kupershmidt 
 or Sawada - Kotera equation. However his transformation cannot be used to our equations (4) 
 because the Kawamoto transformation breaks the time transformation. Indeed notice that 
 our function $r$ is zero dimensional and our time transformation (27) preserve weigth in contrast to 
 the Kawamoto transformation. 
   
\section{Generalization of the Lax Operator}

We have already commented in section {\bf 2} on the fact that the
second order Lax operator for the system of equations in
(\ref{harrydym}) can be generalized and leads to a new third order
equation (\ref{thirdorder}). Under a hodograph transformation
(\ref{transformation}), it can be shown that this new equation
(\ref{thirdorder}) maps into a linear equation of the form
\begin{equation}
p_{\tau} = p_{yyy}.\label{linear1}
\end{equation}

In this section, we will study further generalizations of such
Lax operators.  
First, let us consider the most general Lax operator of third order
parametrized by only one zero-dimensional function $u$ of the form
\begin{equation}
L= u^3\partial_{x}^{3} + k_1 u^2 u_{x}\partial_{x}^{2} + (k_2 u^{2}
u_{xx} + k_3 u u_x^2)\partial_x + 
(k_4 u^{2} u_{xxx}+k_5 u u_{x} u_{xx}+k_6 u_x^3),
\end{equation}
where $k_i,i=1..6$ are arbitrary constant coefficients. It can be
checked that, at the level of fifth order equations, the Lax equation
\begin{equation} 
\frac{\partial L}{\partial t} = \left[\left(L^{5/3}\right)_{\geq 2},L
  \right], 
\end{equation}
leads to consistent equations only for four choices of the
coefficients $ k_i$. For the two cases where $k_1=0$ or
$k_1=\frac{3}{2}$ and all other coefficients
vanishing, we obtain the same fifth order equation as in
(\ref{newequations}), as we have already pointed out in section {\bf 2}. On
the other hand, the Lax operator with $k_{1}=3$ with all other
coefficients vanishing,
\begin{equation}
L= u^3\partial_{x}^{3} + 3 u^{2} u_x \partial_{x}^{2},
\end{equation}
leads to the equation
\begin{equation}
u_t= u^{5} u_{5x} + 5 u u_{x} u_{4x} + \frac{5}{2} u^{4} u_{xx} u_{xxx}
+ \frac{15}{4} u^{3} u_{x}^{2} u_{xxx}. \label{third}
\end{equation}
(Parenthetically we remark here that under a hodograph transformation this
Lax operator goes over to the one for the Sawada-Kotera equation.)
Finally, for $k_{1}=3, k_{2}=k_{3}=1$, with all other coefficients
vanishing, the Lax operator
\begin{equation}
L= u^3\partial_{x}^{3} + 3 u^{2} u_x\partial_{x}^{2} + (u^{2} u_{xx} +
u u_x^2)\partial_x = (u\partial)^3 ,
\end{equation}
yields the equation
\begin{eqnarray}
u_t &=& \big (u^{5} u_{4x} + 5 u^{4} u_{x} u_{3x} + 5 u^{4} u_{xx}^2  +
    5 u^{3} u_{x}^{2} u_{xx} \big )_x \nonumber\\  \label{fourth} 
\end{eqnarray}
Under a hodograph transformation, it can be shown that with
appropriate Miura transformations, equation (\ref{third}) can be
mapped to either Kaup-Kupershmidt or Sawada-Kotera equations, much
like the earlier case discussed. However, under a hodograph
transformation, interestingly enough, the highly nonlinear equation
(\ref{fourth}) maps into a linear equation of the form,
\begin{equation}
p_{\tau} =p_{5y}.\label{linear2}
\end{equation}

In a similar manner, one can carry out the general analysis of the
fourth order Lax operator parameterized by only one variable. It is,
of course, much more involved and we simply quote the results
here. There are four such Lax operators that lead to consistent
seventh order equations following from
\begin{equation}
\frac{\partial L}{\partial t} = \left[\left(L^{7/4}\right)_{\geq 2},
  L\right].
\end{equation}
Three of these with
\begin{eqnarray}
L_{1} & = & u^{4} \partial_{x}^{4} + 6 u^{3} u_{x} \partial_{x}^{3} +
\left(6 u^{2} u_{x}^{2} + 2 u^{3} u_{xx}\right)
\partial_{xx},\nonumber\\
\noalign{\vskip 4pt}%
L_{2} & = & u^{4} \partial_{x}^{4} + 4 u^{3} u_{x} \partial_{x}^{3} +
2\left( u^{2} u_{x}^{2} + u^{3} u_{xx}\right) \partial_{x}^{2} =
\left(u^{2} \partial_{x}^{2}\right)^{2},\nonumber\\
\noalign{\vskip 4pt}%
L_{3} & = & u^{4} \partial_{x}^{4} + 2 u^{3} u_{x} \partial_{x}^{3},
\end{eqnarray}
lead to the same dynamical equation (whose form is complicated and we
do not write it here). On the other hand,
\begin{eqnarray}
L_{4} & = & u^{4} \partial_{x}^{4} + 6 u^{3} u_{x} \partial_{x}^{3} +
\left(4 u^{3} u_{xx} + 7 u^{2} u_{x}^{2}\right)
\partial_{x}^{2}\nonumber\\
\noalign{\vskip 4pt}%
 &  & \quad + \left(u^{3} u_{xxx} + 4 u^{2} u_{x} u_{xx} + u
u_{x}^{3}\right) \partial_{x} = (u\partial)^4,
\end{eqnarray}
leads to a very different equation that is highly nonlinear
\begin{eqnarray}
u_t &=& \big (u_{6x}u^7 + 14u_{5x}u_xu^6 + 28u_{4x}u_{xx}u^6 + 56u_{4x}u_x^2u^5+ \nonumber \\
&& 14u_{3x}^2u^6 + 
 168u_{3x}u_{xx}u_xu^5 + 70 u_{3x}u_x^3u^4 + 42u_{xx}^3u^5 + \nonumber \\
 && 126u_{xx}^2u_x^2u^4+21u_{xx}u_x^4u^3 \big)_x  
\end{eqnarray}
However,
under a hodograph transformation (\ref{transformation}), this equation
goes over to a seventh order linear equation of the form
\begin{equation}
p_{\tau} = p_{7y}.\label{linear3}
\end{equation}
Equations (\ref{linear1}), (\ref{linear2}) and (\ref{linear3}) are
quite puzzling and even more so because of the fact that under the
hodograph transformation, the Lax operators go over respectively to
\begin{equation}
L^{(2)}\rightarrow \partial_{y}^{2},\quad L_{4}^{(3)}\rightarrow
\partial_{y}^{3},\quad L_{4}^{(4)}\rightarrow \partial_{y}^{4}.
\end{equation}
%
All these equations interestingly follows from the same Lax operator $L=u\partial$. Indeed 
this Lax operator constitue the following hierarchy of equations 
\begin{equation}
\frac{\partial L}{\partial t_n} = [L^{n+1}_{\geq 2}, L]  \label{hier}
\end{equation}
where $L=u\partial$ and $n=1,2,3,..$. For $n=1$ we obtain 
\begin{equation}
u_{t_1} = u^2u_{xx}
\end{equation} 
while  for n=2 we obtain the 
equation (7) . For $n=4$ and for $n=6$    we have equations (49) and (54) respectively.
Interestingly all these equations belongs to the 
lineralizable hierarchy of nonlinear partial differential equations considered by M. Euler , N. Euler and 
N. Petersson  \cite{euler1,euler2}. It is possible to write these equations in terms of the following 
recursion operator  \cite{euler1}
\begin{equation}
u_{t_n} = R^{n-1} u_{t_1}
\end{equation}
where $n=1,2,3,...$ and 
\begin{equation}
R = u\partial + u^2u_{xx} \partial^{-1}u^{-2}. \label{rec}
\end{equation}
In order to prove that this recursion operator generates the whole
hierarchy  of equations 
(\ref{hier}) we extract this operator directly from the Lax operator
using a generalization of the method by 
Gurses, Karasu and Sokolov \cite{karas}. Let us note that 
\begin{equation}
L^{n+1}_{\geq 2} = (L(L^{(n-1)+1})_{\geq 2})_{\geq 2} + (L(L^n)_{ < 2})_{\geq 2} = 
L(L^{(n-1)+1})_{\geq 2} + S,
\end{equation}
where $L=u\partial$ in the present case and it follows that
$S=a\partial^2$ with  $a$  an arbitray function to be determined. 
Substituting this  into (\ref{hier}) we obtain  
\begin{equation}
\frac{\partial L}{\partial t_n} = L\frac{\partial L}{\partial t_{n-1}} + au_{xx}\partial +(2au_x-
ua_x)\partial^2, \label{row}
\end{equation}
which determines 
\begin{equation}
a=u^2 \partial^{-1} \frac{u_{t_{n-1}}}{u^2}. 
\end{equation}
With this equation (\ref{row}) leads to 
\begin{equation}
u_{t_{n}} = R u_{t_{n-1}},
\end{equation}
where $R$ coincides with  (\ref{rec}) for $n>1$.

\section{Conclusion}

In this paper, we have studied the properties of a nonlinearly
dispersive integrable system of fifth order and its associated
hierarchy. We have described a Lax representation for such a system
which leads to two infinite series of conserved charges and two
hierarchies of equations that share the same conserved charges. We
have constructed two compatible Hamiltonian structures as well as
their Casimir functionals. One of the structures has a single Casimir
functional while the other has two. This  allows us to extend the
flows into negative order and clarifies the meaning of two different
hierarchies of positive flows. We have studied the behavior of these
systems under a hodograph transformation and have shown that they are
related to the Kaup-Kupershmidt and the Sawada-Kotera
equations under appropriate Miura transformations. We have also
discussed briefly some properties associated with the generalization
of second, third and fourth order Lax operators.

\section*{Acknowledgment}

This work was supported in part by US DOE grant number
DE-FG-02-91ER40685 as well as by NSF-INT-0089589.

\end{document}